\documentstyle[pra,tighten,aps,epsfig,multicol]{revtex} 
\begin{document} 
\title{Generalised coherent states for $SU(n)$ systems}
\author{Kae Nemoto} 
\address{Centre for Laser Science, Department of Physics,\\ 
The University of Queensland, 
QLD 4072 Australia} 
\date{\today} 
\maketitle

\begin{abstract}
Generalised coherent states are developed for $SU(n)$ systems for arbitrary $n$.  
This is done by first iteratively determining explicit representations 
for the $SU(n)$ coherent states, and then determining parametric representations 
useful for applications.  For $SU(n)$, the set of coherent states is isomorphic to a coset
space $SU(n)/SU(n-1)$, and thus 
shows the geometrical structure of the coset space.  These results provide a convenient 
$(2n - 1)$--dimensional space for the description of arbitrary $SU(n)$ systems.  
We further obtain the metric and measure on the coset space, and 
show some properties of the $SU(n)$ coherent states.
\end{abstract}

\section{Introduction} \label{intro}
Coherent states were originally constructed and developed for the Heisenberg--Weyl group 
to investigate quantized electromagnetic radiation \cite{boson}.  These 
coherent states were generated by the action of the Heisenberg--Weyl group operators on the
vacuum state which led to group theoretic generalizations by Peleromov \cite{peleromov_book}
and Gilmore \cite{ann_phys_74_gilmore}.  These two mathematical frameworks differ 
in some points, such as the representations of groups and the reference states; these 
differences are summarized in \cite{rev_mod_phys_62_zhang}.  Coherent states for $SU(2)$ 
have also been called atomic coherent states \cite{j_math_a_4_radcliffe,phys_rev_a_6_arecchi}, 
and have been found useful for treating atom systems, and also for investigations of 
quantum optical models such as nonlinear rotators \cite{barry89}.  
$SU(2)$ coherent states have been successfully applied to the analysis
of the classical limit of quantum systems, and more recently, to the 
investigations of nonlinear quantum systems and quantum entanglement 
\cite{j_phys_b_32_horak,phys_rev_a_57_brif}.

In spite of these many successful $SU(2)$ coherent state applications, 
not much work has been done towards generalising the analysis to other $SU(n)$ groups, 
although $SU(3)$ symmetries were employed to treat a schematic nuclear shell 
model \cite{leboeuf}.  
More recently, this lack has been addressed for $SU(3)$ systems with the explicit
construction of the $SU(3)$ coherent states \cite{j_phys_a_31_guntzmann,pre_nemoto}, the
calculation of Clebsch--Gordon coefficients \cite{byrd_physics_9803029,j_math_phys_38_rowe}, and
the investigation of Wigner functions \cite{j_math_phys_40_rowe}.  Further, geometrical phases
for $SU(3)$ systems have been discussed in \cite{ann_phys_253_khanna}.  
These developments for $SU(3)$ are technologically useful and allow treatment of more 
complex quantum systems such as coupled Bose--Einstein condensates 
\cite{phys_rev_a_sjuly_nemoto}.

In this paper, we construct a set of explicit coherent states for
$SU(n)$, and apply group theoretic techniques to facilitate the 
investigation of nonlinear quantum systems and quantum entanglement.  
In order to construct explicit coherent states, we
need to specify the group representation and the reference states.  
For the chosen group representation, it is necessary to show a useful decomposition and
a parameterization giving usable expressions for the coherent states.  Formal approaches 
to the definition of coherent states are often not readily applicable.  For instance, 
while the Baker--Campbell--Hausdorff relation derived for 
$SU(n)$ \cite{j_math_a_30_weigert}
can be used to define coherent states, this approach does not yield explicit
formulae and parameterisations.  

In this paper, we employ the decomposition for $SU(n)$ in
\cite{j_math_phys_40_rowe} and exploit its symmetric parameterisation.  A set of
coherent states of $SU(n)$ is called an orbit, and is produced by the action of 
group elements on a reference state which here is chosen to be the highest weight state.  
For instance, for $SU(2)$ the highest weight state for a spin {\it 1/2} system is 
spin-up, and the orbit is the surface of a 3-sphere.  For general $n$, this orbit
corresponds to a $(2n-1)$--sphere, which is isomorphic to the coset space
$SU(n)/SU(n-1)$.  The geometrical properties of this coset space 
generalise the $SU(3)$ properties described in \cite{j_math_phys_6_beg}.  
The coset space considered here, $SU(n)/SU(n-1)$, differs slightly from the 
coset space normally considered for coherent states, $SU(n)/U(n-1)$, by
including an arbitrary phase.  
The coset space $SU(n)/SU(n-1)$ enables us to provide a more general method 
to construct coherent states.  Developing the representations and decompositions of 
higher rank groups becomes rapidly messy, however 
the decomposition in \cite{j_math_phys_40_rowe} leads to a systematic 
procedure for the derivation of the coherent states on the coset space 
$SU(n)/SU(n-1)$ without additional complexity.  
Thence we can easily extract an arbitrary phase carrying no physical significance 
for application to physical systems.

In Section \ref{sec-decom}, we obtain an iterative equation in $SU(n)$ coherent states for the simplest
irreducible unitary representation of $SU(n)$.  We also show the geometrical 
structure of the coset space $SU(n)/SU(n-1)$, and provide the metric and 
measure on the space.
In Section \ref{arbitrary-rep}, our analysis is generalised to the case of finding coherent
states of irreducible unitary representations for arbitrarily large dimension, 
and parametric representations are derived. 
We also show some properties of the coherent states.  Finally, we summarize our results 
in Section \ref{summary}.

\section{Decomposition and coset spaces for fundamental representations 
of $SU(\lowercase{n})$} \label{sec-decom}
In order to construct the $SU(n)$ coherent states for the fundamental 
$n \times n$ matrix representation, we first specify the reference state $|\phi_0\rangle$ 
as $(1,0,\ldots,0)^T$, where $T$ denotes transpose.  
This state is a highest weight state, in the sense that it is annihilated by each
of the $SU(n)$ raising operators.  
The raising (lowering) operators $J^h_j$ are equivalent to elementary matrices $e^h_j, \; h<j$
($h>j$) in the $n\times n$ matrix representation.  Appendix A shows the commutation relations
of these matrices.  In this section we review the construction of $SU(2)$ coherent states, 
which provides the origin of the recursive relation of the $SU(n)$ coherent states.  We 
then derive the displacement operators for $SU(3)$ and $SU(4)$, 
employing the $n\times n$
matrix representation of \cite{j_math_phys_40_rowe}.  Finally our results are extended to the 
$SU(n)$ case.

\subsection{Review of $SU(2)$} \label{sec-su2}
Elements $g \in SU(2)$ in the fundamental $2 \times 2$ matrix representation of $SU(2)$ 
may be parameterized as
\begin{equation}   \label{su2 matrix}
  g(\theta, \varphi_1, \varphi_2) = 
  \left( \begin{array}{cc}
  	    e^{i\varphi_1} \cos{\theta} & -e^{-i\varphi_2} \sin{\theta} \\
	    e^{i\varphi_2} \sin{\theta} & e^{-i\varphi_1} \cos{\theta} \\
	 \end{array} \right),
\end{equation}
where angles are real and lie on $0\leq \theta \leq \pi /2$ and $0\leq \varphi_1$, 
$\varphi_2 \leq 2\pi$ respectively.  
The standard approach to obtain a set of coherent states corresponding to $SU(2)/SU(1)$ 
begins with the decomposition of this matrix as
\begin{equation} \label{a-phase}
  g(\theta^{\prime}, \varphi_1, \varphi_2) = 
  \left( \begin{array}{cc}
  	    \cos\frac{\theta^{\prime}}{2} & -e^{-i(\varphi_2-\varphi_1)} \sin\frac{\theta^{\prime}}{2} \\
	    e^{i(\varphi_2-\varphi_1)} \sin\frac{\theta^{\prime}}{2} & \cos\frac{\theta^{\prime}}{2} \\
	 \end{array} \right)
  \left( \begin{array}{cc}
  	    e^{i\varphi_1} & 0 \\
	    0 & e^{-i\varphi_1}\\
	 \end{array} \right),
\end{equation}
where the new variable $\theta^{\prime}$ is introduced as $\theta^{\prime}=2\theta$.  
The action of the right matrix on the highest weight vector 
$|\phi_0\rangle = \left( \begin{array}{c}  
				1\\ 
				0
			 \end{array}\right)$
changes only a phase factor and does not otherwise change the highest weight vector.  
Absolute phases $\varphi_1$ and $\varphi_2$ by themselves do not carry physical 
significance, and only their difference is physically important.  This allows the removal of 
an arbitrary phase usually done by setting $e^{i\varphi_1}=1$.  
Now the action of the left matrix on the highest weight state gives us coherent states 
$|{\mathbf n}_2^{\prime}\rangle = g|\phi_0\rangle = 
			\left( \begin{array}{c}  
				\cos\frac{\theta^{\prime}}{2}\\ 
				e^{i\varphi_2} \sin\frac{\theta^{\prime}}{2}\\
			 \end{array}\right),$
which corresponds to the 2--sphere, i.e. the surface of three--dimensional ball, with the unit vector
$(\cos\theta^{\prime},e^{i\varphi_2}\sin\theta^{\prime})$, or 
$(\cos\theta^{\prime},\sin\theta^{\prime}\cos\varphi_2,
\sin\theta^{\prime}\cos\varphi_2)$ in real coordinates, and the measure 
$d\mu_2^{\prime}= \sin{\theta^{\prime}}d\theta^{\prime} d\varphi_2$ \cite{peleromov_book}.
However this method is not very convenient to construct coherent states for general $SU(n)$. 
The decomposition of $SU(n)$ equivalent to (\ref{a-phase}) is not trivial especially for larger
$n$, and is dependent on the choice of which arbitrary phase is extracted.
In this paper, to avoid using the equivalent decomposition to
(\ref{a-phase}), we begin more generally with the parameterisation (\ref{su2 matrix}), 
derive coherent states and then easily remove an arbitrary phase from our
$SU(n)$ coherent states.  

We now apply $g(\theta, \varphi_1,\varphi_2)$ of (\ref{su2 matrix}) to the highest weight 
state $|\phi_0\rangle$.  
The action yields 
the $SU(2)$ coherent states 
\begin{equation} \label{su_2 cs}
|{\mathbf n}_2\rangle = g|\phi_0\rangle = 
			\left( \begin{array}{c}  
				e^{i\varphi_1} \cos{\theta}\\ 
				e^{i\varphi_2} \sin{\theta}\\
			 \end{array}\right),
\end{equation}
which correspond to points on a 3--sphere 
with unit vector 
$(e^{i\varphi_1} \cos{\theta}, e^{i\varphi_2} \sin{\theta})$, 
from which we derive the expression for the metric on the sphere
\begin{equation}
  |ds_2|^2 = d\theta^2 + \cos^2({\theta})d\varphi_1^2 + \sin^2({\theta})d\varphi_2^2,
\end{equation}
and the measure associated with this metric \cite{j_math_phys_6_beg} as 
\begin{equation} \label{mu2}
  d\mu_2 = \cos{\theta}\sin{\theta} d\theta d\varphi_1 d\varphi_2.
\end{equation}
From this set of coherent states, we now give the procedure to obtain the $SU(2)$ 
coherent states $|{\mathbf n}_2^{\prime}\rangle$.  This is done by setting $e^{i\varphi_1}=1$ 
and introducing the variable $\theta^{\prime}$ so the coherent states now correspond to 
2--sphere.  This shows that one can readily remove an arbitrary phase from our $SU(n)$ 
coherent states without changing the decomposition of the group representation.

For the convenience of the later use, we here introduce $\lambda$--matrices and their
parameterisation of $SU(2)$.
$g$ may also be parameterised with using   
$\lambda$--matrices (i.e. Pauli matrices) 
$\lambda_1  = \sigma_1 =
  \left( \begin{array}{cc}
  	    0  &  1  \\
	    1  &  0\\
	 \end{array} \right)$, 
$\lambda_2  = \sigma_2 =
  \left( \begin{array}{cc}
  	    0  &  -i  \\
	    i  &  0\\
	 \end{array} \right)$, 
and 
$\lambda_3  =  \sigma_3 =
  \left( \begin{array}{cc}
  	    1  &  0  \\
	    0  &  -1\\
	 \end{array} \right)$, as 
\begin{equation}
  g=e^{i\alpha \lambda_3}e^{i\beta \lambda_2}e^{i\gamma \lambda_3},
\end{equation}
where $\varphi_1=\alpha+\gamma$, $\varphi_2=-\alpha+\gamma$, and $\theta=-\beta$, viz 
$\alpha = \frac{1}{2}(\varphi_1-\varphi_2)$, $\beta=-\theta$, and
$\gamma=\frac{1}{2}(\varphi_1+\varphi_2)$.

\subsection{Structure of $SU(n)$ for arbitrary $n$} \label{sec-sun}

We here employ the symmetric parameterisation for the $SU(n)$ matrices provided in
\cite{j_math_phys_40_rowe} to obtain an iterative equation for the $SU(n)$ coherent states.  
This matrix representation efficiently yields the orbit of the highest weight state 
$(1,0,\cdots,0)^T$, because of its symmetric decomposition.  The parameterisation
influences the structure of the iterative equation, which we demonstrate by example 
for small $n$.  We derive firstly the $SU(3)$ coherent states (which may be compared with
the simplest case of \cite{pre_nemoto}), and secondly the $SU(4)$ coherent states.  
For each example, the expression for the coherent
states shows their geometrical structure and determines the metric
and measure of the coset space isomorphic to the coherent states.  
These examples are then generalised to $SU(n)$ by determining the iterative 
equation for the $SU(n)$ coherent states.  Lastly, we 
give the measure of the coset space $SU(n)/SU(n-1)$.

An arbitrary element $g \in SU(n)$ in the $n \times n$ matrix representation
\cite{j_math_phys_40_rowe} may be parameterised as 
\begin{eqnarray}
  g &=& 
  \left( \begin{array}{cccc}
  	   1 & 0  & \cdots & 0\\
	   0 & \; & \; & \;\\
	   \vdots & \; & X_{n-1} & \;\\
	   0 & \; & \; & \;
	 \end{array}\right)
  \left( \begin{array}{cccc}
  	    e^{i\varphi} \cos{\theta} & \, & -\sin{\theta} \\
	    \sin{\theta} & \, & e^{-i\varphi} \cos{\theta} & 0\\
	    \;\\
	    \; & 0 & \; & I_{n-2}\\
	 \end{array} \right)
  \left( \begin{array}{cccc}
  	   1 & 0  & \cdots & 0\\
	   0 & \; & \; & \;\\
	   \vdots & \; & Y_{n-1} & \;\\
	   0 & \; & \; & \;
	 \end{array}\right)\\
\label{su_n matrix}  
&=& L_{n-1} M(\theta, \varphi) R_{n-1}
\end{eqnarray}
where $X_{n-1}, Y_{n-1}$ are the appropriate $(n-1)\times (n-1)$ matrices representing elements
of $SU(n-1)$, and $I_{k}$ is the $k \times k$ 
identity matrix, and we have defined three matrices $L_{n-1}$, $M$, $R_{n-1}$ for convenience.  
(See Appendix B for a justification of this parameterisation.)

\subsubsection{Structure of $SU(3)$} \label{sec-su3}
For $SU(3)$, since the matrices $X_{n-1}, Y_{n-1}$ may be parameterised as (\ref{su2 matrix}), 
(\ref{su_n matrix}) gives 
\begin{eqnarray}
\lefteqn{g(\varphi_1,\xi_1,\varphi_2,\varphi,\theta,\varphi_3,\xi_2,\varphi_4)}
\nonumber\\
  & = &
  \left( \begin{array}{ccc}
  	    1  &  0  &  0\\
  	    0  &  e^{i\varphi_1} \cos{\xi_1} & -e^{-i\varphi_2} \sin{\xi_1} \\
	    0  &  e^{i\varphi_2} \sin{\xi_1} & e^{-i\varphi_1} \cos{\xi_1} \\
	 \end{array} \right)
  \left( \begin{array}{ccc}
  	    e^{i\varphi} \cos{\theta} & -\sin{\theta} & 0\\
	    \sin{\theta} & e^{-i\varphi} \cos{\theta} & 0\\
	    0 & 0 & 1\\
	 \end{array} \right) \nonumber\\
& &  \left( \begin{array}{ccc}
  	    1  &  0  &  0\\
  	    0  &  e^{i\varphi_3} \cos{\xi_2} & -e^{-i\varphi_4} \sin{\xi_2} \\
	    0  &  e^{i\varphi_4} \sin{\xi_2} & e^{-i\varphi_3} \cos{\xi_2} \\
	 \end{array} \right)\nonumber\\
&=& L_{2} M(\theta, \varphi) R_{2}.
\label{su3 matrix}  
\end{eqnarray}
We take the highest weight state $(1,0,0)^T$ as the reference state
and now obtain the expression of the orbit.  Noting that the right matrix $R_{2}$ does  
not change the reference state, the displacement operator for the $SU(3)$ coherent states 
is the product of the left and middle matrices, $L_{2}M(\theta, \varphi)$.  
The left matrix $L_{2}$
corresponds to $SU(2)$, 
hence the orbit of the reference state is the coset space $SU(3)/SU(2)$.  
The first column of the middle matrix $M(\theta, \varphi)$ and the first and second columns of
the left matrix $L_{2}$ in Eq. (\ref{su3 matrix}) can change the reference state, giving 
\begin{eqnarray}
|{\mathbf n}_3\rangle \equiv g|\phi_0\rangle & = & 
L_{2}
  \left( \begin{array}{c}
  	    e^{i\varphi} \cos{\theta}\\
	    \sin{\theta}\\
	    0\\
  	 \end{array} \right)\nonumber\\
&=&  \left( \begin{array}{ccc}
  	    1  &  0  &  0\\
  	    0  &  0  &  0 \\
	    0  &  0  &  0 \\
	 \end{array} \right)
  \left( \begin{array}{c}
  	    e^{i\varphi} \cos{\theta} \\
	    0\\
	    0\\
  	 \end{array} \right)
+  \left( \begin{array}{ccc}
  	    1  &  0  &  0\\
  	    0  &  e^{i\varphi_1} \cos{\xi_1} & -e^{-i\varphi_2} \sin{\xi_1} \\
	    0  &  e^{i\varphi_2} \sin{\xi_1} & e^{-i\varphi_1} \cos{\xi_1} \\
	 \end{array} \right)
  \left( \begin{array}{c}
  	    0\\
	    \sin{\theta}\\
	    0\\
  	 \end{array} \right)\nonumber\\
&=& \left( \begin{array}{c}
  	    e^{i\varphi} \cos{\theta}\\
	    0\\
	    0\\
  	 \end{array} \right)
+ \sin{\theta}
\left( \begin{array}{c}
  	    0\\
	    \begin{array}[b]{cc}
	    \,\\
	    |{\mathbf n}_2 \rangle
	    \end{array}
  	 \end{array} \right).
\end{eqnarray}
The state $|{\mathbf n}_2\rangle$ is the $SU(2)$ coherent state given in (\ref{su_2 cs}), 
hence the coset space is isomorphic to a 5--sphere which has unit
normal  
\begin{equation}
{\mathbf n}_3 = (e^{i\varphi} \cos{\theta},e^{i\varphi_1}\sin{\theta}\cos{\xi_1},
e^{i\varphi_2}\sin{\theta} \sin{\xi_1}),
\label{su_3 cs}
\end{equation}
metric
\begin{equation}
  |ds_3|^2 = d\theta^2 + \cos^2{\theta}d\varphi^2 + \sin^2{\theta}
  (d\xi_1^2 + \cos^2{\xi_1}d\varphi_1^2 + \sin^2{\xi_1}d\varphi_2^2),
\end{equation}
and measure 
\begin{equation}
d\mu_3 = \cos{\theta} \sin^3{\theta} \cos{\xi_1}\sin{\xi_1}
d\theta d\xi_1 d\varphi d\varphi_1 d\varphi_2.
\end{equation}
We note here that an arbitrary phase in these coherent states for $SU(3)$ 
can be easily removed as discussed in 
Subsection (\ref{sec-su2}), and this process is also applicable to general $SU(n)$ cases.  

$SU(3)$ may also be decomposed using $\lambda$--matrices \cite{ann_phys_253_khanna}, 
which yields a slightly different parameterisation from (\ref{su3 matrix}).  
The $SU(2)$ $\lambda$--matrix decomposition for each matrix in (\ref{su3 matrix}) 
gives the $\lambda$--matrix expression for the $SU(3)$ coherent states.  
The middle matrix of (\ref{su3 matrix}) may be constructed using 
$\lambda_2$ and $\lambda_3$ as
\begin{equation}
M(\theta, \varphi) =   
e^{i\varphi /2 \lambda_3}e^{-i\theta \lambda_2}e^{i\varphi /2 \lambda_3}.
\label{second element}
\end{equation}
The left matrix of (\ref{su3 matrix}) may be expressed by $\lambda_7$, $\lambda_8$, 
and $\lambda_3$. 
For convenience, we define a matrix $\lambda_8^{\prime}$ as $(\sqrt{3}\lambda_8 - \lambda_3)/2$,
which describes the $SU(2)$ diagonal generator $\sigma_3$ in the bottom right corner of the
$SU(3)$ matrix 
$\left( \begin{array}{cc}
1 & 0\\
0 & \sigma_3\\ \end{array} \right)$.  The left matrix is
\begin{eqnarray}
L_{2} & =&  e^{i\alpha \lambda_8^{\prime}}e^{i\beta \lambda_7}e^{i\gamma
\lambda_8^{\prime}}\nonumber\\
& = & e^{i\sqrt{3}\alpha /2 \lambda_8} 
   e^{-i\alpha /2 \lambda_3}
   e^{i\beta \lambda_7}
   e^{i\sqrt{3}\gamma /2 \lambda_8}
   e^{-i\gamma /2 \lambda_3},
\label{first element}
\end{eqnarray}
where $\varphi_1=\alpha+\gamma$, $\varphi_2=-\alpha+\gamma$, and $\xi_1=-\beta$.
These two expressions (\ref{second element}) and (\ref{first element}) give 
\begin{eqnarray}
L_{2}M(\theta, \varphi) & = & 
      e^{i\alpha \lambda_8^{\prime}}e^{i\beta \lambda_7}e^{i\gamma\lambda_8^{\prime}}
      e^{i\varphi /2 \lambda_3}e^{-i\theta \lambda_2}e^{i\varphi /2 \lambda_3} \nonumber\\
& = & e^{i\sqrt{3}\alpha /2 \lambda_8} 
   e^{-i\alpha /2 \lambda_3}
   e^{i\beta \lambda_7}
   e^{i\sqrt{3}\gamma /2 \lambda_8}
   e^{-i\gamma /2 \lambda_3}
  e^{i\varphi /2 \lambda_3}e^{-i\theta \lambda_2}e^{i\varphi /2 \lambda_3}
  \label{su_3 lambda}
\end{eqnarray}
The coherent states $|{\mathbf n}_3\rangle$ in this representation are thus 
\begin{eqnarray}
|{\mathbf n}_3\rangle & & = e^{i\alpha \lambda_8^{\prime}}e^{i\beta \lambda_7}
		e^{i\gamma\lambda_8^{\prime}}
      e^{i\varphi /2 \lambda_3}e^{-i\theta \lambda_2}e^{i\varphi /2 \lambda_3} |\phi_0\rangle
      \nonumber\\
& & =  e^{i\sqrt{3}\alpha /2 \lambda_8} 
   e^{-i\alpha /2 \lambda_3}
   e^{i\beta \lambda_7}
   e^{i\sqrt{3}\gamma /2 \lambda_8}
   e^{-i\gamma /2 \lambda_3}
  e^{i\varphi /2 \lambda_3}e^{-i\theta \lambda_2}e^{i\varphi /2 \lambda_3}|\phi_0\rangle.
\end{eqnarray}

\subsubsection{$SU(4)$ and $SU(n)$ for arbitrary $n$} \label{su4 and}
Next we obtain the $SU(4)$ coherent states by applying the above procedure to the $SU(3)$
coherent states.  This process shows the iterative structure of the $SU(n)$ coherent states,
which allows us to define generalised coherent states for arbitrary $n$.  
An arbitrary element $g\in SU(4)$ can be factored by (\ref{su_n matrix}) as 
\begin{eqnarray}
  g & = & L_{3} M(\theta, \varphi) R_{3}\nonumber \\
  &=& 
  \left( \begin{array}{cccc}
  	   1 & 0  & \cdots & 0\\
	   0 & \; & \; & \;\\
	   \vdots & \; & X_3 & \;\\
	   0 & \; & \; & \;
	 \end{array}\right)
  \left( \begin{array}{cccc}
  	    e^{i\varphi} \cos{\theta} & \, & -\sin{\theta} & 0\\
	    \sin{\theta} & \, & e^{-i\varphi} \cos{\theta} & 0\\
	    \;\\
	    0 & 0 & \; & I_2\\
	 \end{array} \right)
  \left( \begin{array}{cccc}
  	   1 & 0  & \cdots & 0\\
	   0 & \; & \; & \;\\
	   \vdots & \; & Y_3 & \;\\
	   0 & \; & \; & \;
	 \end{array}\right)\\
&=&
  \left( \begin{array}{cc}
         \begin{array}{cc}
  	    1 & 0 \\
	    0 & 1
	 \end{array} &
	 \begin{array}{cc}
	    0 & 0\\
	    0 & 0 
	 \end{array}\\
	 \begin{array}{cc}
	    0 & 0\\
	    0 & 0
	 \end{array} &
	 \begin{array}[t]{c}
	  \; X_2
	 \end{array}
	 \end{array}\right)
  \left( \begin{array}{cccc}
  	    1  &  0  &  0 & 0\\
  	    0  &  e^{i\varphi_1} \cos{\xi_1} & -\sin{\xi_1} & 0\\
	    0  &  \sin{\xi_1} & e^{-i\varphi_1} \cos{\xi_1} & 0\\
	    0  &  0  &  0  &  1\\
	 \end{array} \right)
  \left( \begin{array}{cc}
         \begin{array}{cc}
  	    1 & 0 \\
	    0 & 1
	 \end{array} &
	 \begin{array}{cc}
	    0 & 0\\
	    0 & 0 
	 \end{array}\\
	 \begin{array}{cc}
	    0 & 0\\
	    0 & 0
	 \end{array} &
	 \begin{array}[t]{c}
	  \; Y_2
	 \end{array}
	 \end{array}\right)
	 M(\theta, \varphi) R_{3}, 
\label{su_4 matrix}
\end{eqnarray}
where (\ref{su_n matrix}) has been iteratively applied twice.  
Here $X_3$, $Y_3$ are $SU(3)$ matrices and $X_2$ may be parameterised as  
\begin{equation}
X_2 = 
  \left( \begin{array}{cc}
  	    e^{i\varphi_2} \cos{\xi_2} & -e^{-i\varphi_3} \sin{\xi_2} \\
	    e^{i\varphi_3} \sin{\xi_2} & e^{-i\varphi_2} \cos{\xi_2} \\
	 \end{array} \right).
\end{equation}
Taking the highest weight state $|\phi_0\rangle =(1,0,0,0)^T$ and evaluating 
$g|\phi_0\rangle$ as before, we
observe that only two columns, the first column of $X_3$ and the first column of 
the matrix $M(\theta, \varphi)$, are important.  The $SU(4)$ coherent states are  
\begin{eqnarray}
|{\mathbf n}_4\rangle = g|\phi_0\rangle & = &
  e^{i\varphi} \cos{\theta} \left( \begin{array}{c}
  	    1 \\
	    0 \\
	    0 \\
	    0 \\
	 \end{array} \right)
+  \sin{\theta}\left( \begin{array}{cc}
         \begin{array}{cc}
  	    1 & 0 \\
	    0 & 1
	 \end{array} &
	 \begin{array}{cc}
	    0 & 0\\
	    0 & 0 
	 \end{array}\\
	 \begin{array}{cc}
	    0 & 0\\
	    0 & 0
	 \end{array} &
	 \begin{array}[t]{c}
	  \; X_2
	 \end{array}
	 \end{array}\right)
  \left( \begin{array}{cccc}
  	    1  &  0  &  0 & 0\\
  	    0  &  e^{i\varphi_1} \cos{\xi_1} & -\sin{\xi_1} & 0\\
	    0  &  \sin{\xi_1} & e^{-i\varphi_1} \cos{\xi_1} & 0\\
	    0  &  0  &  0  &  1\\
	 \end{array} \right)
  \left( \begin{array}{c}
  	    0\\
	    1\\
	    0\\
	    0\\
	 \end{array} \right)\nonumber\\
 &=&   \left( \begin{array}{c}
  	    e^{i\varphi} \cos{\theta}\\
	    0\\
	    0\\
	    0\\
	 \end{array} \right)
 +  \sin{\theta} \left( \begin{array}{c}
  	    0  \\
	    \begin{array}{c}
	    \,\\
	    |{\mathbf n}_3\rangle\\
	    \, \end{array}  \end{array} \right).
\end{eqnarray}
We note that the matrix including  
$Y_2$ in (\ref{su_4 matrix}) commutes with the matrix to its right, that is 
$[I_2\otimes Y_2,M(\theta, \varphi)]=0$, 
and does not change the state $|\phi_0\rangle$.
The expression of the metric for this coset
space is 
\begin{equation}
  |ds_4|^2=d\theta^2 + \cos^2({\theta})d\varphi^2 + \sin^2({\theta}) 
  \{d\xi_1^2 + \cos^2({\xi_1})d\varphi_1^2 + \sin^2({\xi_1}) (d\xi_2^2 +
  \cos^2({\xi_2})d\varphi_2^2 + \sin({\xi_2})d\varphi_3)\},
\end{equation}
and the measure is 
\begin{equation}
d\mu_4 =
\cos({\theta})\sin^5({\theta})\cos({\xi_1})\sin^3({\xi_1})\cos({\xi_2})\sin({\xi_2})
\; d\theta \; d\xi_1 d\xi_2 \; d\varphi \; d\varphi_1 d\varphi_2 d\varphi_3.
\end{equation}
We note that the total volume is $(2\pi)^4/(6\cdot 4\cdot 2)$.

This establishes that the $SU(n)$ coherent states $|{\mathbf n}_n\rangle$ in this 
representation may be obtained from the iterative relation
\begin{eqnarray}
|{\mathbf n}_n\rangle = g |\phi_0\rangle &=& 
  \left( \begin{array}{c}
  	    e^{i\varphi} \cos{\theta}  \\
	    0\\
	    \vdots\\
	    0 \\
	 \end{array} \right)
+  \left( \begin{array}{cccc}
  	    1  &  0 & \cdots & 0\\
	    0  \\
	    \vdots& \; & X_{n-1}\\
	    0 \\
	 \end{array} \right)
  \left( \begin{array}{c}
  	    0\\
	    \sin{\theta}\\
	    0\\
	    \vdots\\
	    0\\
	 \end{array} \right)\nonumber\\
&=&   \left( \begin{array}{c}
  	    e^{i\varphi} \cos{\theta}\\
	    0\\
	    \vdots\\
	    0\\
	 \end{array} \right)
 +  \sin{\theta} \left( \begin{array}{c}
  	    0  \\
	    \begin{array}{c}
	    \,\\
	    |{\mathbf n}_{n-1}\rangle\\
	    \, \end{array}\\
	 \end{array} \right),
\end{eqnarray}
where $X_{n-1}$ are $SU(n-1)$ matrices, and $|{\mathbf n}_{n-1}\rangle$ is an $SU(n-1)$ 
coherent state. 
Since $|{\mathbf n}_n\rangle$ is the unit vector of the $(2n-1)$--sphere, 
the measure on the hypersphere is 
\begin{equation}
d\mu_n=\cos{\theta}\sin^{2n-3}({\theta})\cos({\xi_1})\sin^{2n-5}({\xi_1})\cdots
\cos({\xi_{n-2}})\sin({\xi_{n-2}}) \; 
d\theta \; d\xi_1 \cdots d\xi_{n-2} \; d\varphi \; d\varphi_1 \cdots
d\varphi_{n-1}.
\end{equation}

\section{Arbitrary $SU(\lowercase{n})$ representations} \label{arbitrary-rep}
We extend the results of the previous section to irreducible unitary 
representations of arbitrarily large dimension for $SU(n)$.  
We define infinitesimal operators and the basis 
of the group representation.  Using the decomposition of $SU(n)$, we derive an iterative 
equation for the $SU(n)$ coherent states and further obtain its recurrence equation.  For 
the purpose of applications, some properties of the $SU(n)$ coherent states are given.
 
\subsection{Infinitesimal operators} \label{infin-op}
We denote $T^N_n$ as a representation of $SU(n)$ where the size number $N$ determines
the dimension of the representation.  A set 
of simultaneous normalised eigenstates of the Cartan operators $J^h_h$ $(1<h \leq n-1)$ is
employed as an appropriate basis to describe the set of coherent states.  
This basis will be
denoted as $|m_1,\ldots, m_n\rangle$ where the $m_j$ satisfy $N=\sum^n_{j=1}m_j$.  
The basis elements are also simultaneous
eigenstates of the size operator $\hat{N}$ such that 
$\hat{N}|m_1,\ldots, m_n\rangle = N|m_1,\ldots, m_n\rangle$.  
For $SU(2)$, these are equivalent to the angular momentum eigenstates.  The $n^2-n$
raising operators $J^h_j$, $1\leq h<j\leq n$, and the same number of lowering operators 
$J^h_j$, $1\leq j<h\leq n$, of $SU(n)$ satisfy the relations 

$[$raising operators, $(h<j)]$
\begin{equation}
J^h_j|m_1,\ldots,m_h,\ldots,m_j,\ldots,m_n\rangle
= \sqrt{(m_h+1)m_j} \; |m_1,\ldots,m_h+1,\ldots,m_j-1,\ldots,m_n\rangle,
\label{raising op}
\end{equation}
$[$lowering operators, $(h>j)]$
\begin{equation}
J^h_j|m_1,\ldots,m_j,\ldots,m_h,\ldots,m_n\rangle
= \sqrt{m_h(m_j+1)} \; |m_1,\ldots,m_h-1,\ldots,m_j+1,\ldots,m_n\rangle.
\end{equation}
For Cartan operators $J^h_h$, $1\leq h\leq n-1$ we have 
\begin{equation}
J^h_h|m_1,\cdots,m_h,\cdots,m_j,\cdots,m_n\rangle
= \sqrt{\frac{2}{h(h+1)}}\left( \sum^h_{k=1}m_k - hm_{h+1}\right)
 |m_1,\cdots,m_h,\cdots,m_j,\cdots,m_n\rangle.
\end{equation}

\subsection{$SU(n)$ coherent states} \label{sec-sun-co-state}

It is appropriate to choose $|\phi_0\rangle = |N,0,\cdots , 0\rangle$ 
as the highest weight state, since the action of any raising operator (\ref{raising op}) on
this state gives zero. 
The parameterization (\ref{su_n matrix}) shows that the representation $T^N_n(g)$ 
may be decomposed as
\begin{equation}
T^N_n(g)=T^N(L_{n-1})T^N(M)T^N(R_{n-1}).
\end{equation}
The action of the representation $T^N(R_{n-1})$ does not change the highest weight state as we have 
seen in the examples in the previous section, hence the coherent state is determined as  
\begin{equation}
|{\mathbf n}^N_n\rangle =T^N(L_{n-1})T^N(M(\theta, \varphi))|N,0,\ldots,0\rangle.
\label{decom g}
\end{equation}
The right element $T^N(M(\theta, \varphi))$ in (\ref{decom g}) acts as an $SU(2)$ 
operator on the subspace 
$|m_1,m_2\rangle$, a cross section obtained by taking the first two elements of 
$|m_1,m_2,\ldots,m_n\rangle$.  
It is well-known \cite{peleromov_book} that an arbitrary $g \in SU(2)$ may be decomposed as 
\begin{eqnarray}
g &=& e^{-\zeta^{\ast} J^1_2}e^{-\nu J^1_1}e^{\zeta J^2_1}e^{i\varphi_1 J^1_1}\nonumber \\
&=& e^{\zeta J^2_1}e^{\nu J^1_1}e^{-\zeta^{\ast} J^1_2}e^{i\varphi_1 J^1_1}.
\label{su_2 decom}
\end{eqnarray}
The parameters in the above expressions correspond to the angle parameters 
in (\ref{su2 matrix}) as $\zeta = e^{i(\varphi_2-\varphi_1)}\tan{\theta}$, and $\nu=\ln{\cos{\theta}}$.
Setting $\varphi_1=\varphi$ and $\varphi_2=0$, $T^N(M(\theta, \varphi))$ 
is decomposed 
as (\ref{su_2 decom}), and acts on the subspace $|N,0\rangle$ of $|N,0,\cdots,0\rangle$ as 
\begin{equation}
T^N(M(\theta, \varphi))|N,0,\ldots,0\rangle
 = \sum^{N}_{j=0}e^{i\varphi(N-j)}\sin^j({\theta})\cos^{N-j}({\theta})
\left( \begin{array}{c}
	N\\
	j\\
	\end{array}\right)
^{1/2}|N-j,j,0,\ldots,0\rangle 
\end{equation}
The left element of the decomposition, $T^N(L_{n-1})$, does not change the first element of the
state $|N-j,j,0,\ldots,0\rangle$, and acts on the subspace $|j,0,\ldots,0\rangle$ in the state 
$|N-j\rangle \otimes |j,0,\ldots,0\rangle$.
This element acts as an $SU(n-1)$ operator on the subspace $|j,0,\ldots,0\rangle$, which 
generates the $SU(n-1)$ coherent states, giving 
\begin{equation}
|{\mathbf n}^N_n\rangle = \sum^{N}_{j=0}e^{i\varphi(N-j)}\sin^j({\theta})\cos^{N-j}({\theta})
\left( \begin{array}{c}
	N\\
	j\\
	\end{array}\right)
^{1/2}|N-j\rangle \otimes |{\mathbf n}^j_{n-1}\rangle.
\label{su_n cs}
\end{equation}

In order to obtain a more convenient expression of the $SU(n)$ coherent states, we derive a
recurrence relation from the above iterative equation (\ref{su_n cs}).  
The last decomposition in (\ref{su_2 decom}) gives the SU(2) coherent states 
\begin{eqnarray}
|{\mathbf n}_2^N\rangle
 &=& \sum^N_{j=0}e^{ij\varphi_2}e^{i(N-j)\varphi_1}\sin^j({\theta})
\cos^{N-j}({\theta})\left( \begin{array}{c}
	N\\
	j\\
	\end{array}\right)^{1/2}
|N-j,j\rangle \nonumber \\
&=& \sum^N_{j=0}\eta^N_j(\varphi_1,\varphi_2,\theta)|N-j,j\rangle,
\end{eqnarray}
where we define 
\begin{equation}
\eta^N_j(\varphi_1,\varphi_2,\theta) \equiv e^{ij\varphi_2}e^{i(N-j)\varphi_1}\sin^j({\theta})
\cos^{(N-j)}({\theta})\left( \begin{array}{c}
	N\\
	j\\
	\end{array}\right)^{1/2}.
\label{coeff}
\end{equation}
The $SU(3)$ coherent states are constructed using the $SU(2)$ coherent states, and the
relations (\ref{su_n cs}) and (\ref{coeff}) give  
\begin{eqnarray}
|{\mathbf n}^N_3\rangle &=& \sum^{N}_{j_1=0}e^{i\varphi(N-j_1)}\sin^{j_1}(\theta)\cos
^{(N-j_1)}({\theta})
\left( \begin{array}{c}
	N\\
	j_1\\
	\end{array}\right)
^{1/2}|N,j_1\rangle \otimes |{\mathbf n}^{j_1}_2\rangle \nonumber \\
&=&\sum^{N}_{j_1=0}\eta^N_{j_1}(\varphi,0,\theta)\sum^{j_1}_{j_2=0}\eta^{j_1}_{j_2}(\varphi_1,\varphi_2,\xi_1)
|N-j_1,j_1-j_2,j_2\rangle,
\end{eqnarray}
in agreement with the $SU(3)$ coherent states developed in \cite{pre_nemoto}.  

Recursively, the $SU(n)$ coherent states may be expressed by the function 
$\eta^l_k(\alpha, \beta,\gamma)$ of (\ref{coeff}),
\begin{eqnarray} \label{su_co_state}
|{\mathbf n}^N_n\rangle &=& \sum^{N}_{j_1=0}\eta^N_{j_1}(\varphi,0,\theta)
			\sum^{j_1}_{j_2=0}\eta^{j_1}_{j_2}(\varphi_1,0,\xi_1)
			\cdots
			\sum^{j_{n-3}}_{j_{n-2}=0}\eta^{j_{n-3}}_{j_{n-2}}
			(\varphi_{n-3},0,\xi_{n-3})
			\nonumber \\
& & \sum^{j_{n-2}}_{j_{n-1}=0}\eta^{j_{n-2}}_{j_{n-1}}(\varphi_{n-2},\varphi_{n-1},\xi_{n-2})
\; |N-j_1,j_1-j_2, \cdots, j_{n-1}\rangle.
\end{eqnarray}

\subsection{Properties of the $SU(n)$ coherent states} \label{properties}
For the purpose of applications, here we describe some fundamental properties of 
the $SU(n)$ coherent states.  

(1) Stereographic coordinates\\
The decomposition (\ref{su_2 decom}) implies that the $SU(n)$ coherent states may be 
represented in the complex numbers ${\zeta_k}$ such that
$\zeta_k=e^{i(\varphi_{k+1}-\varphi_k)}\tan({\xi_k})$.  
Routine change of variables gives the $SU(n)$ coherent states in this stereographic coordinates
as 
\begin{eqnarray}
|{\mathbf n}^N_n\rangle = e^{i\varphi N} 
 \left(\frac{1}{1+|\zeta|^2}\right)^N
			& \sum^N_{j_1=0} & (\zeta)^{j_1}
			\left( \begin{array}{c}
				N\\
				j_1\\
			\end{array}\right)^{1/2}
			\left(\frac{1}{1+|\zeta_1|^2}\right)^{j_1}\nonumber\\
 			& \sum^{j_1}_{j_2=0} & (\zeta_1)^{j_2}
			\left( \begin{array}{c}
				j_1\\
				j_2\\
			\end{array}\right)^{1/2}
			\left(\frac{1}{1+|\zeta_2|^2}\right)^{j_2}\nonumber\\
			& \vdots & 
			\nonumber\\
 			& \sum^{j_{n-2}}_{j_{n-1}=0} & (\zeta_{n-2})^{j_{n-1}}
			\left( \begin{array}{c}
				j_{n-2}\\
				j_{n-1}\\
			\end{array}\right)^{1/2} 
			|N-j_1,\cdots,j_{n-2}-j_{n-1},j_{n-1}\rangle
\end{eqnarray}
\\
(2) Resolution of unity\\
The set of coherent states provides a resolution of unity in the coset space as
\begin{equation}
\frac{(N+n-1)!}{2\pi^n N!} \int d\mu_n |{\mathbf n}^N_n \rangle \langle {\mathbf n}^N_n| 
= \hat{I}.
\end{equation}
The matrix $|{\mathbf n}^N_n \rangle \langle {\mathbf n}^N_n|$ may be expanded in terms of 
matrices $|N-j_1^{\prime},\ldots,j_{n-1}^{\prime}\rangle\langle N-j_1,\ldots,j_{n-1}|$ by using 
the expansion (\ref{su_co_state}).  The integrals with respect to $\xi_k$ are carried out by
change of integral variables $x\equiv\cos^2(\xi_k)$ allowing 
$\int^{\pi/2}_0 d\xi_k \cos^{(2(m-n)+1)}(\xi_k) \sin^{(2n+1)}(\xi_k)=\frac{1}{2}
\frac{n!(m-n)!}{(m+1)!}$, 
while the integrals in terms of $\varphi_j$ produce delta functions.  
The result of the all integrals
cancels with the normalization factor, and gives us 
$\sum_{j_1=0}^N \cdots \sum_{j_{n-1}=0}^{j_{n-2}} 
|N-j_1,\ldots,j_{n-1}\rangle\langle N-j_1,\ldots,j_{n-1}|=\hat{I}$.  \\
\\
(3) Overlap of two coherent states\\
The overlap of two coherent states may be calculated from (\ref{su_n cs}), as  
\begin{equation}
\langle {\mathbf n^{\prime}}^N_n|{\mathbf n}^N_n\rangle 
= \left(
e^{i(\varphi_{n-1}-\varphi^{\prime}_{n-1})}
	\prod^{n-2}_{k=0}\sin{\xi_k}\sin{\xi^{\prime}_k}
+ \sum^{n-2}_{m=0}\left[ e^{i(\varphi_m-\varphi^{\prime}_m)}\cos{\xi_m}\cos{\xi^{\prime}_m}
\prod^{m-1}_{k=0}\sin{\xi_k}\sin{\xi^{\prime}_k}\right] \right)^N,
\end{equation}
where we have changed the notation of angles, replacing $\theta$ with $\xi_0$, 
and $\varphi$ with $\varphi_0$, and where we have defined  
$\prod^{-1}_{k=0}\sin{\xi_k}\sin{\xi^{\prime}_k}=1$.  

\section{Summary} \label{summary}

In conclusion, we have described $SU(n)$ coherent states for irreducible unitary 
representations for arbitrarily large dimension and some examples for small $n$ demonstrated.  
The geometric structure of the $SU(n)$ coherent states has been represented using 
spherical coordinates.  We also gave expressions for 
the resolution of unity, and the non-orthogonality of the coherent states.  
It was shown the $SU(n)$ coherent states may be recursively derived from $SU(2)$
coherent states.

\acknowledgments{The author would like to thank David De Wit for useful discussions and
comments about group representations and Michael J. Gagen for useful suggestions.  
The author acknowledges the financial support of the Australian 
International Education Foundation (AIEF).}

\appendix
 
\section{$\lambda$--matrices}
In general $SU(n)$ generators can be represented by $n^2-n$ off--diagonal matrices 
and $n-1$ diagonal matrices.  For example, $SU(4)$ has fifteen generators which can be
constructed using twelve off--diagonal matrices and three diagonal matrices \cite{georigi_book}.
We take $\{e^h_j\}$ as a basis
for the group $SU(n)$, where ${e^h_j}$ are elementary matrices.  
We also define $e^h_j, (h<j)$ as raising operators, and $e^h_j, (h>j)$ as
lowering operators respectively.  Non-diagonal elements of this basis are 
\begin{equation}
\{ \beta^h_j=-i (e^h_j-e^j_h), \Theta^h_j= e^h_j+e^j_h, 1 \leq h < j \leq n \}.
\end{equation}
The commutation relations of these non--diagonal elements are  
\begin{equation}
[\beta^h_j, \Theta^k_e ]=-i\delta^k_j\Theta^h_e + i\delta^h_e\Theta^k_j +
i\delta^k_h\Theta^j_e-i\delta^j_e\Theta^k_h.
\end{equation}
The diagonal elements $\{\eta^m_m |1 \leq m \leq n-1\}$ are 
\begin{equation}
\eta^m_m = \sqrt{\frac{2}{m(m+1)}}\left(\sum^m_{j=1}{e^j_j} - m \; e^{m+1}_{m+1}\right).
\end{equation}
For instance in $SU(4)$ the fifteen $\lambda$--matrices are numbered as
\begin{eqnarray}
& &\left\{
\begin{array}{cc}
  \lambda_1 = \Theta^1_2 & \lambda_2 = \beta^1_2 \\
  \lambda_3 = \eta^1_1, \\
\end{array}\right.\nonumber\\
& &\left\{
\begin{array}{cc}
  \lambda_4 = \Theta^1_3 & \lambda_5 = \beta^1_3 \\
  \lambda_6 = \Theta^2_3 & \lambda_7 = \beta^2_3 \\
  \lambda_8 = \eta^2_2, \\
\end{array}\right.\nonumber\\
& &\left\{
\begin{array}{cc}
  \lambda_9 = \Theta^1_4 & \lambda_{10} = \beta^1_4 \\
  \lambda_{11} = \Theta^2_4 & \lambda_{12} = \beta^2_4 \\
  \lambda_{13} = \Theta^3_4 & \lambda_{14} = \beta^3_4 \\
  \lambda_{15} = \eta^3_3.\\
\end{array}\right.
\end{eqnarray}
These $\lambda$--matrices are the generators of the representation $T^1_4$.

These $SU(4)$ generators allow another expression for the coherent states.  
Defining a matrix $\lambda_{15}^{\prime}$ as $(\sqrt{6}\lambda_{15}-\sqrt{3}\lambda_8)/3$, 
the decomposition of $SU(4)$ using $\lambda$--matrices gives expressions for the
coherent states 
\begin{eqnarray}
|{\mathbf n}_4\rangle &=& e^{i\alpha \lambda_{15}^{\prime}}e^{i\beta \lambda_{14}}
		e^{i\gamma\lambda_{15}^{\prime}}
		e^{i\varphi_1 /2 \lambda_8^{\prime}}e^{-i\xi_1 \lambda_7}
		e^{i\varphi_1 /2 \lambda_8^{\prime}} 
      e^{i\varphi /2 \lambda_3}e^{-i\theta \lambda_2}e^{i\varphi /2 \lambda_3} |\phi_0\rangle
      \nonumber\\
&=&	e^{i\sqrt{6}\alpha /3 \lambda_{15}}e^{-i\sqrt{3}\alpha /3 \lambda_8}
	e^{i\beta \lambda_{14}}
	e^{i\sqrt{6}\gamma /3 \lambda_{15}}e^{-i\sqrt{3}\gamma /3 \lambda_8}
	e^{i\sqrt{3}\varphi_1 /4 \lambda_8}e^{-i\varphi_1 /4 \lambda_3}
	e^{-i\xi_1 \lambda_7}
	e^{i\sqrt{3}\varphi_1 /4 \lambda_8}e^{-i\varphi_1 /4 \lambda_3} \nonumber \\
& & 	e^{i\varphi /2 \lambda_3}e^{-i\theta \lambda_2}e^{i\varphi /2 \lambda_3} |\phi_0\rangle,
\end{eqnarray}
where $\varphi_2 = \alpha + \gamma$, $\varphi_3= -\alpha+\gamma$, and $\xi_2=-\beta$.  
These expressions have been obtained directly from (\ref{su_3 lambda}) and (\ref{su_4 matrix}), 
and show the displacement operator for the $SU(4)$ coherent states.

\section{The symmetric parameterization for $SU(\lowercase{n})$} 
We here show a brief proof of the parameterization (\ref{su_n matrix}).  A proof of this
parameterization for $n=3$ was given in \cite{j_math_phys_40_rowe}.  
Showing any element of $g \in SU(n)$ can be transformed into $R_{n-1}^{\dagger}$, we give the
parameterization (\ref{su_n matrix}) as an inverse equation in terms of the element $g$.  
We first review the proof
for $n=3$, and prove (\ref{su_n matrix}) for arbitrary $n$ inductively.  

(i) For n=3, an arbitrary
element $g = 
\left( \begin{array}{ccc}
	x_{11} & x_{12} & x_{13} \\
	x_{21} & x_{22} & x_{23} \\
	x_{31} & x_{32} & x_{33} \end{array}\right)$
 can be transformed as
\begin{equation}
\left( \begin{array}{ccc}
	1 & 0 & 0 \\
	0 & a^* & b^* \\
	0 & -b & a \end{array}\right)
\left( \begin{array}{ccc}
	x_{11} & x_{12} & x_{13} \\
	x_{21} & x_{22} & x_{23} \\
	x_{31} & x_{32} & x_{33} \end{array}\right)
=
\left( \begin{array}{ccc}
	x_{11} & x_{12} & x_{13} \\
	r^3_2 &
	\frac{\sum^3_{k=2}x^*_{k1}x_{k2}}{r^3_2} &
	\frac{\sum^3_{k=2}x^*_{k1}x_{k3}}{r^3_2} \\ 
	0 & \star & \star \end{array}\right),
\label{transform}
\end{equation}
where $a=x_{21}/r^3_2$, $b=x_{31}/r^3_2$ and $r^q_p=\sqrt{\sum^q_{k=p}|x_{k1}|^2}$.  
Applying a matrix $
\left( \begin{array}{ccc}
	x^*_{11} & \sqrt{1-|x_{11}|^2} & 0 \\
	-\sqrt{1-|x_{11}|^2} & x_{11} & 0 \\
	0 & 0 & 0 \end{array}\right)$
from the left on the above matrix (\ref{transform}) gives   
$\left( \begin{array}{ccc}
	1 & 0 & 0 \\
	0 & \star & \star \\
	0 & \star & \star \end{array}\right)$, where 
we used the constraints on $g$, which are $
\sum^3_{j=1}x^*_{jk}x_{jl}=\delta_{kl}$.  With suitably chosen parameters, the inversion of the
above relation gives the devised SU(3) parameterization (\ref{su3 matrix}).
Now we extend this procedure to the general result, and prove it inductively.  

(ii) We assume the result in the case (i), that is, 
for $n=m$ an arbitrary element $g \in SU(m)$ can be parametrized as (\ref{su_n
matrix}), and for any $g$ a matrix $X_{m-1}\in SU(m-1)$ exists such that
\begin{equation}
\left( \begin{array}{cccc}
	1 & 0 & \cdots & 0 \\
	0 & \; & \; & \; \\
	\vdots& \; & X^{\dagger}_{m-1}& \;\\
	0 & \; & \; & \; \end{array}\right) 
\left( \begin{array}{cccc}
	x_{11} & x_{12} & \cdots & x_{1m} \\
	x_{21} & x_{22} & \cdots & x_{2m} \\
	\vdots& \vdots & \vdots & \vdots\\
	x_{m1} & x_{m2} & \cdots & x_{mm} \end{array}\right) 
=
\left( \begin{array}{cccc}
	x_{11} & x_{12} & \cdots & x_{1m} \\
	r^m_2 &
	\frac{\sum^m_{k=2}x^*_{k1}x_{k2}}{r^m_2} & \cdots &
	\frac{\sum^m_{k=2}x^*_{k1}x_{km}}{r^m_2} \\ 
	0 & \star & \star & \star \\
	\vdots & \vdots & \vdots & \vdots \\
	0 & \star & \star & \star \end{array}\right).
\label{assumption}
\end{equation}

(iii) For n=m+1, using (\ref{assumption}), an arbitrary matrix $g \in SU(m+1)$ can be transformed as
\begin{eqnarray}
\left( \begin{array}{cccc}
	I_2 & 0 & \cdots & 0 \\
	0 & \; & \; & \; \\
	\vdots & \; & Y^{\dagger}_{m-1}& \;\\
	0 & \; & \; & \; \end{array}\right) 
\left( \begin{array}{cccc}
	1 & 0 & 0 & \; \\
	0 & \frac{x^*_{21}}{r^{m+1}_2} & 
	\frac{r^{m+1}_3}{r^{m+1}_2}& 0 \\ 
	0 & -\frac{r^{m+1}_3}{r^{m+1}_2} &
	\frac{x_{21}}{r^{m+1}_2}& \;\\ 
	\; & 0 & \; & I_{m-2} \end{array}\right)
\left( \begin{array}{cccc}
	I_2 & 0 & \cdots & 0 \\
	0 & \; & \; & \; \\
	\vdots& \; & X^{\dagger}_{m-1}& \;\\
	0 & \; & \; & \; \end{array}\right) \nonumber\\
\times
\left( \begin{array}{cccc}
	x_{11} & x_{12} & \cdots & x_{1m+1} \\
	x_{21} & x_{22} & \cdots & x_{2m+1} \\
	\vdots& \vdots & \vdots & \vdots\\
	x_{m+11} & x_{m+12} & \cdots & x_{m+1m+1} \end{array}\right)
=
\left( \begin{array}{cccc}
	x_{11} & x_{12} & \cdots & x_{m+1} \\
	r^{m+1}_2 &
	\frac{\sum^{m+1}_{k=2}x^*_{k1}x_{k2}}{r^{m+1}_2} & 
	\cdots& \frac{\sum^{m+1}_{k=2}x^*_{k1}x_{km+1}}{r^{m+1}_2} \\ 
	0 & \star & \cdots & \star\\
	\vdots & \vdots & \vdots & \vdots\\ 
	0 & \star & \cdots & \star \end{array}\right),
\label{m+1}
\end{eqnarray}  
where $I_k$ are $k \times k$ identity matrices.  Using the constraints for $SU(m+1)$ matrices, 
$\sum^{m+1}_{j=1}x^*_{jk}x_{jl}=\delta_{kl}$, the matrix on the right hand side can be
transformed to contain an $SU(m)$ matrix as
\begin{eqnarray}
&\left( \begin{array}{ccc}
	x^*_{11} & \sqrt{1-|x_{11}|^2} & 0 \\
	-\sqrt{1-|x_{11}|^2} & x_{11} & 0 \\
	 0 & \; & I_{m-1} \end{array}\right)&
\left( \begin{array}{cccc}
	x_{11} & x_{12} & \cdots & x_{m+1} \\
	r^{m+1}_2 &
	\frac{\sum^{m+1}_{k=2}x^*_{k1}x_{k2}}{r^{m+1}_2} & 
	\cdots& \frac{\sum^{m+1}_{k=2}x^*_{k1}x_{km+1}}{r^{m+1}_2} \\ 
	0 & \star & \cdots & \star\\
	\vdots & \vdots & \vdots & \vdots\\ 
	0 & \star & \cdots & \star \end{array}\right)\nonumber\\
\lefteqn{=
\left( \begin{array}{cccc}
	1 & 0 & \cdots & 0 \\
	0 & \; & \; & \; \\
	\vdots& \; & Y_{m}& \;\\
	0 & \; & \; & \; \end{array}\right)}
\end{eqnarray}
Since the second matrix can be parametrized equivalently to the $n=m$ case, the product of the
three matrices constructs an $SU(n)$ matrix.  The inversion of this relation gives
\begin{eqnarray}
\lefteqn{
\left( \begin{array}{cccc}
	x_{11} & x_{12} & \cdots & x_{1\, m+1} \\
	x_{21} & x_{22} & \cdots & x_{2\, m+1} \\
	\vdots& \vdots & \vdots & \vdots\\
	x_{m+1\, 1} & x_{m+1\, 2} & \cdots & x_{m+1\, m+1} \end{array}\right)}\nonumber\\
& &=
\left( \begin{array}{cccc}
	1 & 0 & \cdots & 0 \\
	0 & \; & \; & \; \\
	\vdots& \; & X_m & \;\\
	0 & \; & \; & \; \end{array}\right)
\left( \begin{array}{cccc}
	x_{11} & -\sqrt{1-|x_{11}|^2} & \; &  \\
	\sqrt{1-|x_{11}|^2} & x_{11}^* & \; & 0 \\
	\; & \; & \; & \; \\
	\; & 0 & \; & I_{m-1} \end{array}\right)
\left( \begin{array}{cccc}
	1 & 0 & \cdots & 0 \\
	0 & \; & \; & \; \\
	\vdots& \; & Y_m & \;\\
	0 & \; & \; & \; \end{array}\right).
\end{eqnarray}
The elements, $x_{11}$ and $\sqrt{1-|x_{11}|^2}$, can be parametrized as
\begin{equation}
x_{11}=e^{i\varphi}\cos{\theta}, \; \; \sqrt{1-|x_{11}|^2}=\sin{\theta}.
\end{equation}

\end{document}